\documentclass[12pt]{article} 

\usepackage[utf8]{inputenc} 

\usepackage{geometry} 
\usepackage{graphicx, subfig} 

\usepackage{booktabs} 
\usepackage{array} 
\usepackage{paralist} 
\usepackage{verbatim} 
\usepackage{subfig} 

\usepackage{fancyhdr} 
\usepackage{sectsty}
\allsectionsfont{\sffamily\mdseries\upshape} 

\usepackage[nottoc,notlof,notlot]{tocbibind} 
\usepackage[titles,subfigure]{tocloft} 

\usepackage{cite}

\usepackage{amsmath, amsfonts, amsthm, amssymb, amscd}
\usepackage{indentfirst}

\usepackage[pdftex,bookmarksnumbered]{hyperref}

\newtheorem{theorem}{Theorem}

\title{A Note on Computational Complexity of \\ Kill-all Go}
\author{Zhujun Zhang \thanks{E-mail: zhangzhujun1988@163.com} \\Water Bureau of Fengxian District, Shanghai}

\date{November 26, 2019} 

\providecommand{\keywords}[1]{\textbf{\textit{Index terms---}} #1}

\begin{document}
\maketitle

\begin{abstract}

Kill-all Go is a variant of Go in which Black tries to capture all white stones, while White tries to survive.
We consider computational complexity of Kill-all Go with two rulesets, Chinese rules and Japanese rules.
We prove that: (i) Kill-all Go with Chinese rules is PSPACE-hard, and (ii) Kill-all Go with Japanese rules is EXPTIME-complete.

\end{abstract}


\section{Introduction}

Kill-all Go is a live-or-die variant of Go.
The rules of Kill-all Go are as in regular Go, but Black is awarded 17 handicap stones.
If White can obtain at least one living group, he wins, while if Black can prevent this, she wins.

There are many rulesets of Go all around the world, while Chinese rules and Japanese rules are typical rulesets of these.
Usually there is no difference between using Chinese or Japanese rules in practice, however they are different on allowing cycles. 
Cycles of length 2 (termed ko) are forbidden in either of these two rulesets. 
Longer cycles (termed superko) are forbidden in Chinese rules, while longer cycles are allowed in Japanese rules. 
If a board position is repeated, the game has ``no result'' in Japanese rules.

In this note, we discuss computational complexity of Kill-all Go with two rulesets, Chinese rules (superko rule) and Japanese rules (basic ko rule).

\section{Related Work}

Computational complexity of Go and variants of Go was researched in last several decades.
In 1980, Lichtenstein and Sipser \cite{gopspace} considered complexity of Go first, and they proved Go to be PSPACE-hard. 
Their reduction is not sensitive to rulesets, therefore they proved Go with either of Chinese rules or Japanese rules to be PSPACE-hard in fact.
In 1983, Robson \cite{goexptime} proved Go with Japanese rules to be EXPTIME-complete, and an improvement on Robson's proof could be found on Sensei's Library \cite{robsonsproof}.
In 1998, Cr{\^a}{\c{s}}maru \cite{tsumeGo} researched a kind of life and death problem of Go (termed tsumego) in which one of the players has always a unique good move and the other has always only two good moves available to choose from.
Cr{\^a}{\c{s}}maru proved this kind of life and death problem to be NP-complete.
In 2000, Cr{\^a}{\c{s}}maru and Tromp \cite{goladders} considered the ladder which is a technique for capturing stones in Go, and they proved Ladders to be PSPACE-complete.
In 2002, Wolfe \cite{goendgames} proved Go endgames (termed yose) to be PSPACE-hard.
In 2015, Saffidine et al. \cite{atarigo} considered a variant of Go, Atari Go, in which the first capture leads to a victory, and they proved Atari Go to be PSPACE-complete.
They also researched complexity of Phantom Go which is a partially observable variant of Go, and they proved lower and upper bounds for Phantom Go.

Computational complexity of Kill-all Go is still open.
Saffidine et al. \cite{atarigo}  conjectured that Kill-all Go with Chinese rules is PSPACE-hard and Kill-all Go with Japanese rules is EXPTIME-complete.
Moreover, they gave a few hints on this direction, and they suggested using ladders to construct gadgets.
However, using ladders will lead to a large number of vacant points on the game board.
It is difficult to rigorously prove that White can not obtain a living group on these vacant points.
Thus we choose to use Lichtenstein and Sipser's method \cite{gopspace}, and we use capturing races to construct gadgets. 

\section{Complexity of Kill-all Go with Chinese Rules}

\subsection{Overview of the Reduction}

We reduce True Quantified Boolean Formula (TQBF) to Kill-all Go with Chinese rules.
We need to construct start, pipe, white switch, black switch, merge, verify and crossover gadgets.
We construct a capturing race in a start gadget, and the result of the capturing race decides who wins the game.
In a start gadget, White's only hope is to escape through the small breach.
Then he has try to connect the group in the start gadget with a group which has a large number of liberties.
White can not make any eyes in other gadgets, so that The only chance for him to win the game is to obtain a living group in a start gadget.
White switch gadgets are corresponding to choices of universal player in TQBF, while black switch gadgets are corresponding to choices of existential player.
We use two verify gadgets to represent the assignment of each variables in TQBF.
Whether the white group in the start gadget can obtain enough liberties depends on the status of verify gadgets.
We use pipe, merge and crossover gadgets to connect other gadgets according to the formula in TQBF.
In all following gadgets, White is to move.

\subsection{Gadgets}

\begin{figure}[htbp]
	\centering  
	\includegraphics[width=0.48 \linewidth]{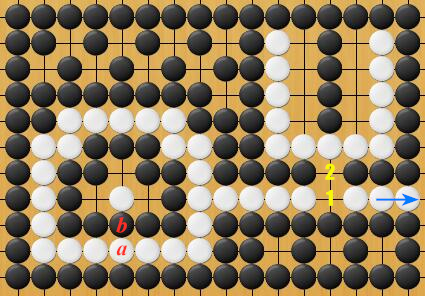}  
	\caption{A start gadget.}  
	\label{start}   
\end{figure}

\textbf{Start gadget.} 
A start gadget for Kill-all Go is illustrated in Figure \ref{start}.
In this gadget, white group $a$ has only one liberty, while black group $b$ has three liberties.
If White connects group $a$ with a group which has a large number of liberties, White will capture group $b$, and he will obtain a living group easily.
While Black must try to prevent this.
Moreover, as we will see later, White can not make any eyes in other gadgets.
The only chance for White to obtain a living group is to try to capture group $b$ in the start gadget.
Thus White has to move at point 1 in order to save group $a$.
Then Black must move at point 2, otherwise, White may move at point 2, and group $a$ will obtain enough liberties so that White can capture group $b$ in at most six moves.
Group $a$ will leave the start gadget through the right exit finally if two players move properly.

\begin{figure}[htbp]
	\centering  
	\includegraphics[width=0.38 \linewidth]{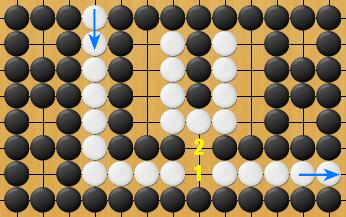}  
	\caption{A pipe gadget.}  
	\label{pipe}   
\end{figure}

\textbf{Pipe gadget.} 
Pipe gadgets are used to connect other gadgets, and a pipe gadget for Kill-all Go is illustrated in Figure \ref{pipe}.
After leaving the start gadget, group $a$ might enter a pipe gadget from north.
Then White should move at point 1, otherwise Black can capture group $a$ immediately.
And then Black must respond at point 2, otherwise group $a$ will obtain seven liberties at least.
Thus group $a$ will leave this pipe gadget through the right exit finally.
Obviously, in a pipe gadget, no black stones will be captured, and White can not make any eyes.
Moreover, we use pipe gadgets to restrict the number of liberties of white groups at each entrance and exit of other gadgets.

\begin{figure}[htbp]
	\centering  
	\includegraphics[width=0.52 \linewidth]{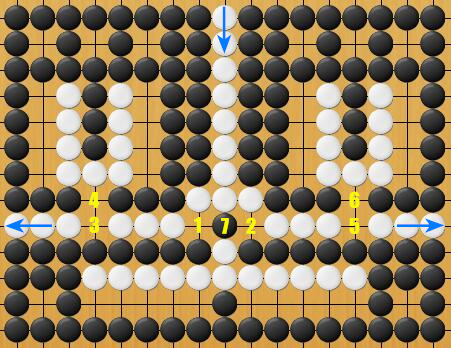}  
	\caption{A white switch gadget.}  
	\label{white_switch}   
\end{figure}

\textbf{White switch gadget.} 
A white switch gadget for Kill-all Go is illustrated in Figure \ref{white_switch}.
When group $a$ enters a white switch gadget from north, White can choose to move at point 1 or point 2, and Black must respond.
For instance, a sequence of proper moves might be: W-1, B-2, W-3, B-4. 
Then group $a$ leaves the gadget through the left exit.
The situation that White chooses to move at point 2 is symmetrical.
If White's first move is not at either point 1 or point 2, Black can capture group $a$.
For instance, the sequence of moves might be: W-3, B-1, W-2, B-5.
If Black does not respond White's first move at point 1 or point 2, group $a$ will obtain a large number of liberties.
For instance, the sequence of moves might be: W-2, B-5, W-1 (captures one black stone at point 7), B-3, W-7.
Note that group $a$ has at most two liberties all along. 
Thus if White tries to capture group $b$ in the start gadget, Black will capture group $a$ first.
Moreover, in a white switch gadget, White has at most one false eye at point 7, while Black can destroy this false eye sooner or later.

\newpage

\begin{figure}[htbp]
	\centering  
	\includegraphics[width=0.55 \linewidth]{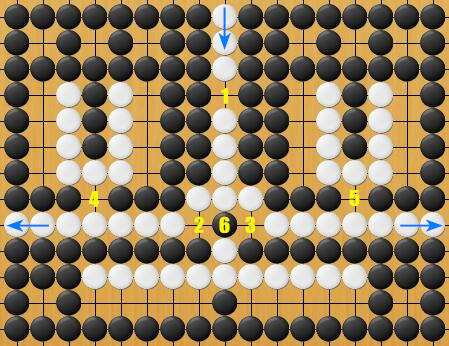}  
	\caption{A black switch gadget.}  
	\label{black_switch}   
\end{figure}

\textbf{Black switch gadget.} 
A black switch gadget for Kill-all Go is illustrated in Figure \ref{black_switch}.
A black switch gadget is similar to a white switch gadget.
When white group $a$ enters a black switch gadget from north, White should move at point 1, otherwise Black can capture group $a$ immediately.
Then Black can choose to move at point 2 or point 3, and White must respond.
For instance, a sequence of proper moves might be: W-1, B-3, W-2, B-4.
Then group $a$ leaves the gadget through the left exit.
The situation that Black chooses to move at point 2 is symmetrical.
If Black does not move at either point 2 or point 3, group $a$ will obtain a large number of liberties.
For instance, the sequence of moves might be: W-1, B-4, W-3, B-5, W-2 (captures one black stone at point 6), B-at any legal point (note that Black can not move at point 6 to capture white stones since pipe gadgets provide liberties), W-6.
Similarly, in a black switch gadget, group $a$ has at most two liberties all along. If White tries to capture group $b$, Black will capture group $a$ first.
Moreover, White has at most one false eye at point 6.

\begin{figure}[htbp]
	\centering  
	\includegraphics[width=0.55 \linewidth]{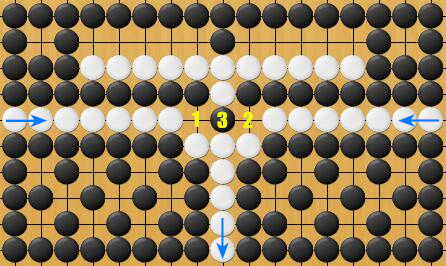}  
	\caption{A merge gadget.}  
	\label{merge}   
\end{figure}

\textbf{Merge gadget.} 
A merge gadget for Kill-all Go is illustrated in Figure \ref{merge}.
If group $a$ enters this gadget from west, White should move at point 1, and then Black must respond at point 2.
If Black does not respond, White may move at point 2 to capture one black stone at point 3.
Since pipe gadgets provide liberties, Black can not move at point 3 to capture white stones.
Thus White can move at point 3 next, so that group $a$ will obtain a large number of liberties.
The situation that group $a$ enters the gadget from east is symmetrical.
Group $a$ will leave the start gadget through the lower exit finally if two players move properly.
Moreover, White has at most one false eye at point 3 in a merge gadget.

\newpage

\begin{figure}[htbp]
	\centering  
	\includegraphics[width=0.5 \linewidth]{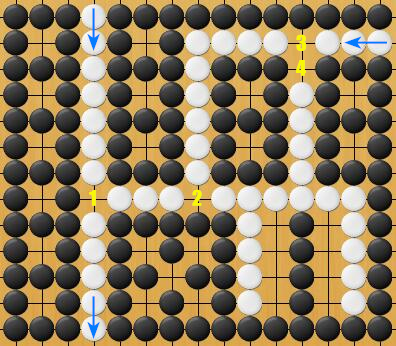}  
	\caption{A verify gadget.}  
	\label{verify}   
\end{figure}

\textbf{Verify gadget.} 
A verify gadget for Kill-all Go is illustrated in Figure \ref{verify}.
When group $a$ enters this gadget from north, White should move at point 1, otherwise Black can move at point 1 to capture group $a$ immediately.
Then Black must respond at point 2, otherwise White can move at point 2 so that group $a$ will obtain a large number of liberties.
When group $a$ enters this gadget from east, there are two cases: 
(i) If there is one black stone on point 2. 
Black may capture group $a$ by moving at point 4 after White moves at point 3; 
(ii) If there is no stones on point 2.
White may move at point 3, and then he can choose between point 2 or point 4 in next move, so that Black fails to prevent group $a$ from obtaining a large number of liberties.
Moreover, White can not make any eyes in a verify gadget.

\begin{figure}[htbp]
	\centering  
	\includegraphics[width=0.5 \linewidth]{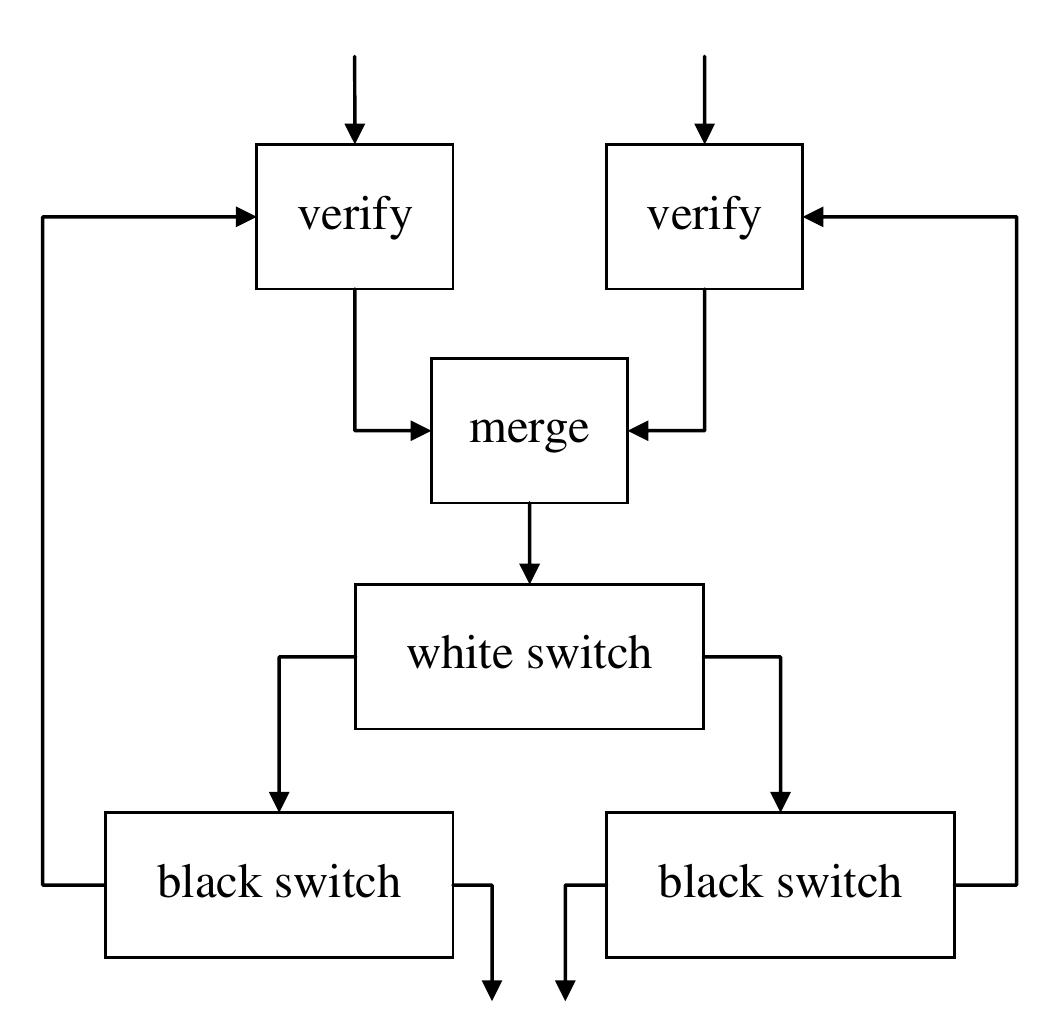}  
	\caption{A crossover gadget.}  
	\label{crossover}   
\end{figure}

\textbf{Crossover gadget.} 
A crossover gadget is composed of above gadgets, and it is illustrated in Figure \ref{crossover}.
When group $a$ enters the left verify gadget from north, it traverses a verify gadget, a merge gadget and a white switch gadget continuously. 
Then White has to make group $a$ enter the right black switch gadget, otherwise Black may make group $a$ enter the left verify gadget so that group $a$ will be captured.
In the right black switch gadget, Black can not make group $a$ enter the right verify gadget.
Otherwise group $a$ will obtain a large number of liberties, since it did not ever traverse the right verify gadget.
The situation that group $a$ enters the right verify gadget from north is symmetrical.
Note that, once group $a$ traverses a crossover gadget through one path, the other path will be locked. 
However, as we will see later, each crossover gadget will never be traversed for more than one time in the instance of Kill-all Go.

\subsection{The Example}

\begin{figure}[htbp]
	\centering  
	\includegraphics[width=0.75 \linewidth]{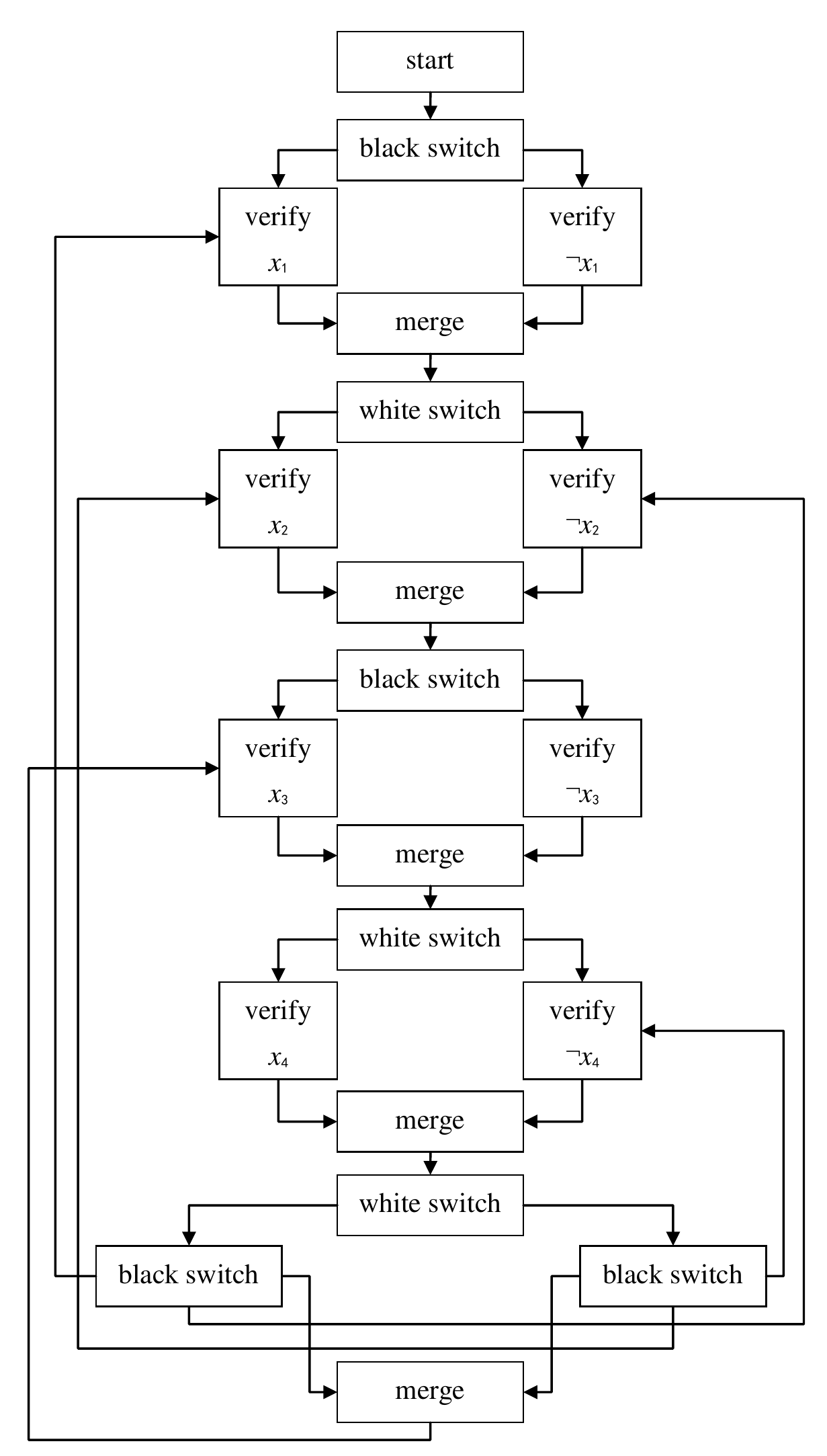}  
	\caption{An instance of Kill-all Go with Chinese rules.}  
	\label{instance1}   
\end{figure}

We use a example to demonstrate how to simulate TQBF by Kill-all Go.
For Boolean formula $\exists x_1 \forall x_2 \exists x_3 \forall x_4 \left( x_1 \vee \neg x_2 \vee x_3 \right) \wedge \left( x_2 \vee x_3 \vee \neg x_4 \right)$, the corresponding instance of Kill-all Go with Chinese rules is illustrated in Figure \ref{instance1}.

In the figure, arrow lines represent pipe gadgets, and each crossroad represents a crossover gadget.
In the instance, black switch gadgets simulate choices of existential player, and white switch gadgets simulate choices of universal player in TQBF.
Thus the escape path of group $a$ is corresponding to assignment of variables $x_1, x_2, x_3$ and $x_4$.
The lower one white switch gadget and two black switch gadgets simulate checking progress.
Thus White chooses the clause to be checked, and Black chooses the literal to be checked.

It is easy to verify that the Boolean formula of TQBF is true if and only if Black can capture all white stones in Kill-all Go.
Since TQBF is PSPACE-complete, we prove the following theorem:

\begin{theorem}
	Kill-all Go with Chinese rules is PSPACE-hard.
\end{theorem}

Actually, we can use same method to prove Kill-all Go with Japanese rules to be PSPACE-hard, however we will obtain stronger result in next section.

\section{Complexity of Kill-all Go with Japanese Rules}

\subsection{Overview of the Reduction}

We use Robson's method \cite{goexptime} to reduce a formula game to Kill-all Go with Japanese rules.
In the formula game, a certain positive Boolean formula is given.
There are two players which are also called White and Black in the formula game, and they move alternately.
In each turn, White changes the assignment of one variable to true, while Black changes the assignment of one variable to false.
A player can not immediately revert the opponent's last move, which is similar to basic ko rule in Go.
If White is to move, and the current assignment makes the formula true, he wins the game.
Robson \cite{kolike} proved this formula game to be EXPTIME-complete.

To simulate the formula game, we also need to construct some gadgets for Kill-all Go with Japanese rules.
Fortunately, the gadgets constructed in last section could be reused here.
Moreover, we need to construct ko gadgets corresponding to variables in the formula game, and we will use a capturing race gadget to replace the start gadget.
There is a large capturing race in a capturing race gadget.
Whether White wins the game depends on whether one of two white groups can obtain enough liberties in a capturing race gadget.
There is one path for each of these two white groups, and each path is connected with ko gadgets.
These two paths are constructed by previous gadgets, and the structures of the paths are corresponding to the formula.
If White can connect one white group with a ko gadget in which he takes the ko, the white group will obtain enough liberties so that White can obtain a living group.

\subsection{Gadgets}

\begin{figure}[htbp]
	\centering  
	\includegraphics[width=0.50 \linewidth]{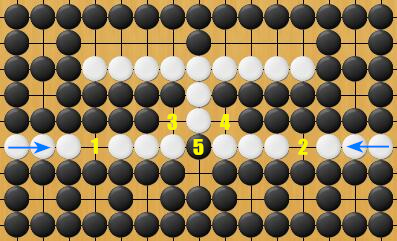}  
	\caption{A ko gadget (Black takes the ko).}  
	\label{ko}   
\end{figure}

\textbf{Ko gadget.} 
A ko gadget for Kill-all Go is illustrated in Figure \ref{ko}.
There is a ko at point 5 in the ko gadget, and each ko gadget is corresponding to a variable in the formula game.
The gadget in which Black takes the ko is corresponding to that assignment of the variable is false, while the gadget in which White takes the ko is corresponding to that assignment of the variable is true.
When a white group enters the gadget, White has to move at point 1 or point 2.
If the ko is taken by Black, she can capture the white group by moving at point 3 or point 5.
If the ko is taken by White, the white group can obtain a large number of liberties, and Black can not prevent this.
For instance, the sequence of moves might be: W-2, B-5, W-4.
Moreover, White has at most one false eye at point 5 in a ko gadget.

\begin{figure}[htbp]
	\centering  
	\includegraphics[width=1 \linewidth]{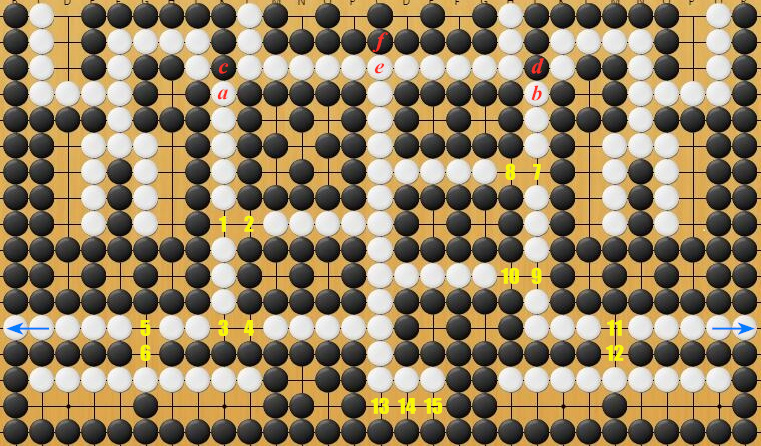}  
	\caption{A capturing race gadget.}  
	\label{capturing_race}   
\end{figure}

\textbf{Capturing race gadget.} 
A capturing race gadget for Kill-all Go is illustrated in Figure \ref{capturing_race}.
There is a large capturing race in this gadget.
Two exits of this gadget are connected with two paths, and each path is connected with ko gadgets according to the formula.
In a capturing race gadget, White's only hope is to capture group $c$ or group $d$.
Once group $c$ or group $d$ is captured, White can make two eyes easily so that he may obtain a living group.
If Black captures groups $a$ and group $b$, White can not prevent Black from connecting group $c$ and group $d$ with living black groups, and White can not make any eyes in other gadgets so that he loses.
If Black captures group $e$, she can connect group $f$ with group $c$ and group $d$, so that White can not obtain any living groups.

In a capturing race gadget, each of group $a$ and group $b$ has only one liberty, and each of group $c$ and group $d$ has three liberties, and group $e$ has seven liberties.
If White immediately tries to capture group $c$ or group $d$, group $a$ or group $b$ will be captured first.
Thus the only way for White to win the game is to connect group $a$ or group $b$ with a group which has a large number of liberties.
Group $a$ or group $b$ should leave the gadget through the left exit or the right exit at the right time, and White should try to connect group $a$ or group $b$ with a ko gadget in which he takes the ko, so that he can capture group $c$ or group $d$.
On the other hand, if Black immediately captures one of group $a$ or group $b$, the other group will leave the gadget through one of two exits, and White may connect it with a ko gadget in which he takes the ko finally.

We discuss strategies of two players in a capturing race gadget in detail:

(1) Suppose that White is to move, and the formula is true under the assignment corresponding to current status of ko gadgets.
Then White may move at point 1, which forces Black to respond at point 2.
If Black does not respond, White can connect group $a$ with group $e$, and these two groups share at least six liberties, so that White can capture group $c$.
After Black moves at point 2, White can move at point 3, which forces Black to move at point 4 similarly.
Next, group $a$ leaves the gadget through the left exit after White moves at point 5.
Since the formula is true, White can connect group $a$ with a ko gadget in which he takes the ko sooner or later, so that group $a$ will obtain enough liberties.
Moreover, group $e$ has at least five liberties, therefore, if Black tries to capture group $e$ in this progress, White will capture group $c$ first.
Thus White can obtain a living group, and he will win the game finally.

(2) Suppose that Black is to move, and the formula is false under the assignment corresponding to current status of ko gadgets.
Then Black may move at point 1 to capture group $a$, and White has to move at point 7, point 9 and point 11 
continuously in order to save group $b$.
After Black responds at point 8, point 10 and point 12, group $b$ leaves the gadget through the right exit.
However, since the formula is false, Black can force group $b$ to enter a ko gadget in which she takes the ko.
Thus White can not prevent Black from capturing group $b$.
Note that, group $d$ has three liberties, while group $b$ has at most two liberties.
If White tries to capture group $d$ in this progress, group $b$ will be captured first.
Thus Black can capture group $a$ and group $b$ sooner or later. 
Since White can not make any eyes in other gadgets, Black can capture all white stones, and she will win the game finally.

(3) Suppose that Black is to move, and the formula is true under the assignment corresponding to current status of ko gadgets.
Then Black must move at a ko gadget to make the formula false, otherwise she will lose the game since she can not prevent group $a$ and group $b$ from leaving the gadget at the same time.
For instance, suppose that Black moves at point 1 or point 2 in order to capture group $a$.
Then group $b$ will obtain a large number of liberties.
Since group $d$ has only three liberties, after White moves at point 7, Black must respond at point 8.
Then White moves at point 9, and Black still has to move at point 10, otherwise, White can connect group $b$ with group $e$, so that these two groups share at least four liberties.
Next, White moves at point 11, and Black responds at point 12.
Group $b$ leaves the gadget through the right exit.
Since the formula is true, group $b$ will enter a ko gadget in which White takes the ko sooner or later.
Note that, if Black tries to capture group $e$ in this progress, group $d$ will be captured first since group $e$ has at least four liberties. 

(4) Suppose that White is to move, and the formula is false under the assignment corresponding to current status of ko gadgets.
Then White must move at a ko gadget to make the formula true, otherwise he will lose the game.
We consider three types of White's improper moves:

(4.1) Suppose that White moves at a ko gadget to connect a ko.
Since the status of ko gadgets is unchanged, Black can capture one of group $a$ and group $b$.
The other white group can not enter a ko gadget in which White takes the ko, even if it can leave the capturing race gadget.

(4.2) Suppose that White moves at point 1 or point 7.
We just discuss the situation that White moves at point 1, since the situation that White moves at point 7 is symmetrical.
Then Black must respond at point 2.
Since the formula is false, group $a$ can not avoid being captured even if it can leave the gadget through the left exit.
If White tries to save group $b$, the sequence of moves might be: W-1, B-2, W-3, B-4, W-5, B-6, ... , W-7, B-8, W-9, B-10.
Note that group $b$ has only one liberty, and group $e$ has only three liberties (point 13, point 14 and point 15) now.
If White moves at point 11 to save group $b$, Black can try to capture group $e$, and White fails to capture group $d$ first.
Moreover, after White moves at point 1, if he moves at a ko gadgets to make the formula true.
Then Black still can move at point 4, so that neither group $a$ or group $b$ can leave the capturing race gadget because of shortage of liberties of group $e$.

(4.3) Suppose that White moves at other points, such as point 2, point 8 or points in other gadgets.
This situation is similar to case (4.1), and Black may capture one of group $a$ and group $b$ immediately.
Then the other white group can not enter a ko gadget in which White takes the ko.

\subsection{The Example}

\begin{figure}[htbp]
	\centering  
	\includegraphics[width=1 \linewidth]{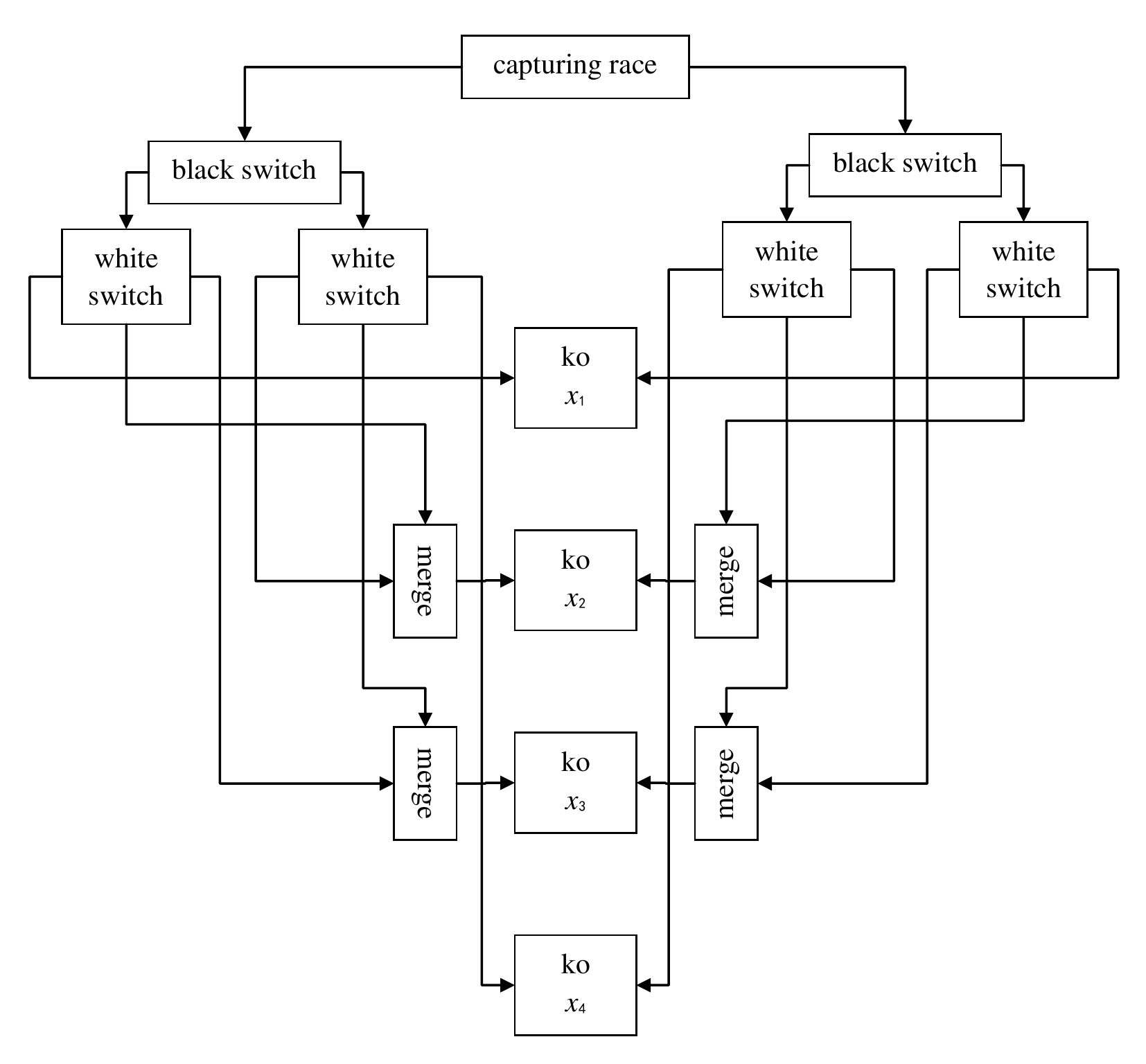}  
	\caption{An instance of Kill-all Go with Japanese rules.}  
	\label{instance2}   
\end{figure}

We also use a example to demonstrate how to simulate the formula game by Kill-all Go.
For formula $\left( x_1 \vee x_2 \vee x_3 \right) \wedge \left( x_2 \vee x_3 \vee x_4 \right)$, the corresponding instance of Kill-all Go with Japanese rules is illustrated in Figure \ref{instance2}.

Two exits of the capturing race gadget are connected with two symmetrical paths which are constructed according to the formula.
Group $a$ and group $b$ in the capturing race gadget can enter ko gadgets through these two paths.
Suppose that the formula is true under the current assignment, White can connect group $a$ or group $b$ with a ko gadget in which he takes the ko.
Suppose that the formula is false under the current assignment, Black can force group $a$ or group $b$ to enter a ko gadget in which she takes the ko.
Thus each of the two players has to move at the ko gadgets until her or his opponent makes a mistake, and the basic ko rule should be obeyed, which simulates the formula game.

It is easy to verify that White wins the formula game if and only if White can obtain at least one living group in Kill-all Go.
Moreover, since the number of board positions of Go is exponential, Kill-all Go with Japanese rules is in EXPTIME.
Thus we prove the following theorem:

\begin{theorem}
	Kill-all Go with Japanese rules is EXPTIME-complete.
\end{theorem}

\section{Conclusion}

We reduce TQBF to Kill-all Go with Chinese rules, so that we prove Kill-all Go with Chinese rules to be PSPACE-hard.
We reduce a formula game to Kill-all Go with Japanese rules, so that we prove Kill-all Go with Japanese rules to be EXPTIME-complete.

The instances of Kill-all Go constructed in this note could be regarded as huge life and death problems which are local skill problems of Go usually.
The instance constructed in Section 4 could be regarded as a huge superko which is a interesting subject of Go.

The open problem is computational complexity of Kill-all Go with Chinese rules.
Since Kill-all Go with Chinese rules is in EXPSPACE, it might be PSPACE-complete, EXPTIME-complete or even EXPSPACE-complete.
Robson \cite{expspaceproblem} introduced a “no-repeat” version of a formula game, and he proved the new formula game to be EXPSPACE-complete.
The no-repeat rule of the formula game is similar to superko rule in Go, so we can try to reduce the no-repeat formula game to Go or Kill-all variant of Go.


\bibliographystyle{plain}
\bibliography{ref}

\end{document}